\documentstyle[12pt]{article}
\def\k{{\rm {\bf k}}}
\def\p{{\rm {\bf p}}}
\def\q{{\rm {\bf q}}}

\begin{document}

\hfill BI-TP 95/06

\hfill February 1995

\vspace{1.5cm}

\begin{center}
{\bf THE HIGH TEMPERATURE DISPERSION EQUATION FOR LONGITUDINAL PLASMA
OSCILLATIONS IN TAG }
\footnote{Research is partially supported by "Volkswagen-Stiftung"}

\vspace{1.5cm}
{\bf O.K.Kalashnikov}
\footnote{Permanent address: Department of
Theoretical Physics, P.N.Lebedev Physical Institute, Russian
Academy of Sciences, 117924 Moscow, Russia. E-mail address:
kalash@td.lpi.ac.ru}

Fakult\"at f\"ur Physik

Universit\"at Bielefeld

D-33501 Bielefeld, Germany

\vspace{2.5cm}
{\bf Abstract}
\end{center}
The calculations in the temporal axial gauge (TAG)
are revised and a new prescription is introduced to avoid the well-known
TAG-singularity. With this prescription we use the TAG-formalism to
calculate the one-loop dispersion equation for the longitudinal plasma
oscillations in the high temperature limit and find the complete
selfconsistency of TAG for pragmatic aims. Our result reproduces the
earlier known dispersion equation obtained in covariant gauges and
this equality  explicitly demonstrates the gauge independence
of the dispersion law in the high temperature limit and its reliability.

\newpage

\section{Introduction}

The resent results obtained for the Debye mass screening beyond the leading
term [1-6] revived the
interest to the selfconsistency of calculations made in an axial gauge,
namely in the temporal axial gauge (TAG). Today (bearing in mind the need
to have results beyond the perturbative expansions) it is a very actual problem
since all the axial gauges due to the simple Slavnov-Taylor identities (like
in QED) are singled out and very convenient when any nonperturbative
calculations are performed. Using these gauges there is a chance to build
explicitly the nonperturbative vertex function in accordance with the
corrections made for the dressed gluon propagator. In doing in this manner
one has an opportunity to keep explicitly the gauge covariance throughout all
calculations and thus solving the selfconsistency problem of any
nonperturbative approaches beyond the standard expansions. TAG where the gauge
vector is parallel to the medium vector $u_\mu$ has the additional advantage
in selecting the tensor structures which determine the Green function
formalism and their amount is minimum in TAG. The latter is the main argument
of doing TAG (between other axial gauge) to be more preferable for practice
although it has one very essential defect: a pole at $p_4=0$. There
were many attempts to eliminate this singularity [7-10]
but all these attempts being or a rather
complicated or very radical destroy all advantages found in TAG. But it
is also well-known that the problem concerns (in any case on the one-loop
level, but see also [11]) only the two sums which
can be easily redefined to be correct [12,13]. Both these sums are well-known
and have a rather simple form
\setcounter{equation}{0}
\begin{eqnarray}
T\sum_{p_4}1/p_4=0\,,\qquad      T\sum_{p_4}1/p_4^2=0
\end{eqnarray}
where $p_4=2\pi nT$ with $ n=0,\pm 1,\pm 2 $ and so on. Our goal is to show
that TAG (after the new additional prescription is made) is a reliable gauge
for any pragmatic aims and to repeat the important result (earlier found only
in covariant gauges): the equation which defines the longitudinal plasma
oscillations in the high temperature limit. In doing so we can check explicitly
the gauge-independence  of the high temperature dispersion equation for the
plasma ocsillations (there is no damping in this limit) and reliability of TAG.

\section{The TAG formalism and the exact $\Pi_{44}$-function}

As it was mentioned above the choice of the gauge vector
$n_\mu$  to be parallel to the medium one $u_\mu$ considerably simplifies
at the beginning all the Green function technique. The exact polarization
tensor (for TAG) is determined by two scalar functions only [12]
\begin{eqnarray}
\Pi_{\mu\nu}(k)=G\left(\delta_{\mu\nu}-\frac{k_\mu k_\nu}{k^2}\right)+
(F-G)B_{\mu\nu}
\end{eqnarray}
and due to this fact the gluon propagator has a rather simple form
\begin{eqnarray}
{\cal D}_{ij}(k)=
\frac{1}{k^2+G}\left(\delta_{ij}-\frac{k_ik_j}{\k^2}\right)+
\frac{1}{k^2+F}\frac{k^2}{k_4^2}\frac{k_ik_j}{\k^2}
\end{eqnarray}
The scalar functions $F(k)$ and $G(k)$ are defined as
follows
\begin{eqnarray}
G(k)=\frac{1}{2}\left(\sum_{i}\Pi_{ii}+
\frac{k_4^2}{\k^2}\Pi_{44}\right)\,,\qquad
F(k)=\frac{k^2}{\k^2}\Pi_{44}
\end{eqnarray}
and they should be calculated through the graph (or another)
representation for $\Pi$. Due to a peculiarity of the temporal
axial gauge the ${\cal D}_{44}(k)$-function is completely
eliminated from  the formalism as well as the ${\cal D}_{4i}(k)$ and
${\cal D}_{i4}(k)$-ones. The exact Slavnov-Taylor identity for
the $\Gamma_{3}$-vertex function has the same form as in QED
\begin{eqnarray}
r_\mu\Gamma_{\mu\nu\gamma}^{abc}(r,\,p,\,q)=
igf^{abc}[{\cal D}_{\nu\gamma}^{-1}(p)-{\cal D}_{\nu\gamma}^{-1}(q)]
\end{eqnarray}
and namely this fact is doubtless an advantage of the axial gauge.

The exact graph representation for the gluon polarization tensor
is well-known (see e.g. [12,13]) and contains (for any axial gauge)
the standard four nonperturbative graphs. However there is one important
simplification if one considers the $\Pi_{44}$-function only; in this
case two "one-loop" nonperturbative graphs give at once its exact expression
since the remaining graphs (the two very complicated ones) are equal to zero
at the beginning (the unique situation which takes place only in TAG).
Thus after some algebra being performed the exact expression for the
$\Pi_{44}$-function is found to be
\begin{eqnarray}
&&\Pi_{44}(q_4,|\q|)=
\frac{g^2N}{\beta}\sum_{p_4}
\int\frac{d^3\p}{(2\pi)^3}{\cal D}_{ii}(p)-\\
&&\hspace{1cm}-\frac{g^2N}{2\beta}\sum_{p_4}
\int\frac{d^3\p}{(2\pi)^3}
(2p+q)_{4}\left[{\cal D}_{li}(p+q)
\Gamma_{ij4}(p+q,\, -p,\, -q){\cal D}_{jl}(p)\right]\nonumber
\end{eqnarray}
and we are going to calculate its leading order in the high temperature limit.

For calculating in the lowest order it is enough to put in Eq.(6) only the
bare vertex which has a rather simple form
\begin{eqnarray}
\Gamma_{4ji}^{abc}(q,r,p)=-igf^{abc}\delta_{ij}(r_4-p_4)
\end{eqnarray}
and when all algebra being performed one can obtain the more simple expression
for $\Pi_{44}(q_4,|\q|)$ which now is valid only on the one-loop level. This
expression is found to be
\begin{eqnarray}
\Pi_{44}(q_4,|\q|)&=&
\frac{g^2N}{\beta}\sum_{p_4}
\int\frac{d^3\p}{(2\pi)^3}{\cal D}_{ii}^{(0)}(p)-\\
&-&\frac{g^2N}{2\beta}\sum_{p_4}
\int\frac{d^3\p}{(2\pi)^3}
(2p+q)_{4}^2\left[{\cal D}_{li}^{(0)}(p+q)
{\cal D}_{il}^{(0)}(p)\right]\nonumber
\end{eqnarray}
where ${\cal D}_{ij}^{(0)}(p)$ is the standard bare propagator for
TAG
\begin{eqnarray}
{\cal D}_{ij}^{(0)}(p)=\frac{1}{p^2}(\delta_{ij}+\frac{p_ip_j}{p_4^2})
\end{eqnarray}

\section{ The high temperature limit of the longitudinal plasma oscillation in
the leading order}

Now a simple algebra should be performed within Eq.(8). Then we calculate a few
sums and namely here we encounter the main TAG disadvantage: the well-known
pole which should be redefined. After doing this we obtain a rather complicated
expression which will be calculated in the so-called high-temperature limit.
This is a special manner of calculation which reproduced only $T^2$-terms
but in a straightforward way.

Thus after all algebra being performed within Eq.(8) it can be put into
the form
\begin{eqnarray}
\Pi_{44}(q_4,|\q|)&=&\frac{g^2N}{\beta}\sum_{p_4}\int\frac{d^3p}{(2\pi)^3}
\left\{\left(\frac{2}{p^2}+\frac{1}{p_4^2}\right)\right.\\
&-&\left.\frac{1}{2}(2p+q)_4^2
\left[\frac{1+cos^2(\p+\q|\p)}{p^2(p+q)^2}
+\frac{cos^2(\p+\q|\p)}{p_4^2(p+q)_4^2}\right.\right.\nonumber\\
&+&\left.\left.(1-cos^2(\p+\q|\p))\left(\frac{1}{p^2(p+q)_4^2}
+\frac{1}{(p+q)^2p_4^2}\right)\right]\right\}\nonumber
\end{eqnarray}
where a new abreviation is introduced to be
\begin{eqnarray}
cos^2(\p+\q|\p)=\frac{(\p+\q|\p)^2}{(\p+\q)^2\p^2}
\end{eqnarray}
In what follows we calculate only the leading $T^2$-term within Eq.(10).
For this case (i.e. when the $T^2$-term is only kept ) the two last terms
in Eq.(10) can be at once omitted since in the
high temperature region due to the factor $(1-cos^2(\p+\q|\p))$ the
$T^2$-term is absent in their asymptotic behaviour. One can check this
explicitly by taking into account the regularized value for the appropriate
sum which is found to be
\begin{eqnarray}
\frac{1}{\beta}\sum_{p_4}\frac{1}{p_4^2[(q+p)_4^2+\omega^2]}=
-\frac{(\omega^2-q_4^2)n^{(B)}(\omega)}{\omega(\omega^2+q_4^2)^2}
\end{eqnarray}
where $n^{(B)}(\omega)=[exp(\beta\omega)-1]^{-1}$ and then calculating the
asymptotic behaviour for the integral over momenta in accordance with (10).
To find Eq.(12) we use the new ansatz which is proposed here for
treating any integral with a $p_4$ singularity. This ansatz is very close
to the well-known Landshoff's $\alpha$-prescription [7] but it is essentially
modified due to an exponential factor. More precisely it has the form
\begin{eqnarray}
\big[\frac{1}{p_4^2}\big]\longrightarrow
\raisebox{-2.5mm}{$\buildrel{\displaystyle\lim}\over{\scriptstyle\alpha\to 0}$}
\frac{{\displaystyle exp[-\frac{\alpha^2 T^2}{|p_4|^4}]}}
{p_4^2+\alpha^2}\,,\qquad \alpha>0
\end{eqnarray}
where we use (and it is very essential) the absolute value of $p_4$ in the
exponential factor. Exploiting this new ansatz one can easily check Eq.(1) and
we use it as well to calculate all the sums in Eq.(10). In particular we find
that
\begin{eqnarray}
\frac{1}{\beta}\sum_{p_4}\frac{(2p+q)_4^2}{p_4^2(p+q)_4^2}=0
\end{eqnarray}
and bearing in mind Eq.(14) the last term from the second line in Eq.(10) can
be
omitted as well. So, we arrive at the complicated (but well-defined) sum which
being treated as usual has the form
\begin{eqnarray}
I&=&\frac{1}{\beta}\sum_{p_4}\int\frac{d^3p}{(2\pi)^3}
\frac{(2p+q)_4^2}{p^2(p+q)^2}\nonumber\\
&=&\int\frac{d^3p}{(2\pi)^3}\frac{n^{(B)}(p)}{2|\p|}\left\{\big[
\frac{(2i|\p|+q_4)^2}{(i|\p|+q_4)^2+|\p+\q|^2}+{h.c.}\big]\right.\nonumber\\
%% FOLLOWING LINE CANNOT BE BROKEN BEFORE 80 CHAR
&+&\left.\big[\frac{(2i|\p|+q_4)^2}{(i|\p|+q_4)^2+|\p-\q|^2}+{h.c.}\big]\right\}
\end{eqnarray}
All other sums in Eq.(10) are trivial and now using Eq.(15) we can directly
calculate the $T^2$-term within Eq.(10) which reproduces the final result for
this task. In doing so we introduce the dimensionless variable $z=|\p|/T$ (see
the expression for $n^{(B)}(p)$ in Eq.(15)) and keep only the leading term
which
due to dimension of the integral in Eq.(15) is the $T^2$-term. All other terms
of this expansion reproduce the next-to-leading order terms and should be
omitted. Using this ansatz the integral over $|\p|$ is calculated exactly and
only the integral over the angular variables still remains to compare with the
known expression. The final result has the form
\begin{eqnarray}
\Pi_{44}(q_4,|\q|)&=&\frac{g^2T^2N}{6}\left\{1-(3q_4^2+\q^2)\right.\nonumber\\
&\times&\left.\int\limits_0^1\frac{dx}{q_4^2+\q^2x^2}+2q_4^2(\q^2+q_4^2)
\int\limits_0^1\frac{dx}{(q_4^2+\q^2x^2)^2}
\right\}
\end{eqnarray}
and that is completely the same as  previously known in the covariant gauges
(see [14,15] and also [16]) Then doing as usual (with $\xi=\frac
{\omega}{|\p|})$ one obtains that
\begin{eqnarray}
\omega_{||}^2(\xi)=\xi^2p_{||}^2(\xi) \, ,\qquad   1<\xi<\infty
\end{eqnarray}
where
\begin{eqnarray}
p_{||}^2(\xi)=3\omega_{pl}^2\left\{\frac{\xi}{2}log[\frac{\xi+1}{\xi-1}]-1
\right\}
\end{eqnarray}
and we can conclude that the entire (for any momenta) dispersion curve of the
longitudinal plasma oscillations is gauge-independent, of course, if the
leading
$T^2$-term is kept only (here $\omega_{pl}^2=g^2T^2N/6)$).

\section{Conclusion}

To summarize we have proposed here a new prescription  which successfully
treats
all TAG-singularities and completly restores the reliability of the
TAG-formalism for pragmatic aims. To prove this we demonstrate
TAG-possibility by reproducing with
a new prescription the high temperature dispersion equation for longitudinal
plasma oscillations and find that these calculations are reliable and equal
to the well-known results. Of course, from a theoretical point of view
there is a number of unsolved questions (e.g. the periodicity problem) but
in as much as TAG is not a rigorous gauge at the beginning (and despite on that
we choose it) all other theoretical questions can be plainly forgotten. Thus
gathering all well-known results
one can conclude that there are no problems with applying TAG on the one-loop
level and it is not clear today indeed such problems arise for multi-loop
(nonperturbative) calculations where TAG has at once a number of
the serious advantages. However it is also known that the next-to-leading
order Debye mass found in TAG [3,4]  has a power screening
behaviour which is in contradiction with that obtained in covariant gauges
( [1,2] and [5,6] ). Of course, it is a serious problem, but
there are no reasons to consider that the discrepancy found is a
defect of TAG where all calculations are selfconsistent and do not suffer
from any infrared divergencies. We should notice that this is not the case
for calculations made in other gauges ( [1,2] and[5,6] )  which
are very sensitive to the infrared cutoff which is usually associated with the
so-called gluon magnetic mass. Nevertheless it is doubtful
that the strong infrared sensitivity of the next-to-leading order Debye mass
indeed should be taken place (although see [17]) and it is not excluded that
namely this fact leads to a discrepancy with the calculations made in the
different gauges.

\newpage

\begin{center}
{\bf Acknowledgements}
\end {center}

I would like to thank Rudolf Baier for useful discussions as well as
all the colleagues from the Department of Theoretical Physics of the
Bielefeld University for the kind hospitality.

\begin{center}
{\bf Appendix}
\end{center}

Here we compare the calculations made with the different prescriptions (namely
with the prescription proposed here and with another one (see Ref.[10] ) for
the sums of Eq.(1). Our calculation concerns the well-known sum
\begin{eqnarray}
T\sum_{p_4}1/p_4^2=0
\end{eqnarray}
which should be equal zero to reproduce explicitly many well-known results
found in other gauges. With the prescription proposed here this sum is found to
be
\begin{eqnarray}
\frac{1}{\beta}\sum_{p4}\frac{1}{p_4^2}\longrightarrow
\raisebox{-2.5mm}{$\buildrel{\displaystyle\lim}\over{\scriptstyle\alpha\to 0}$}
\frac{1}{\beta}\sum_{p4}\frac{{\displaystyle exp[-\frac{\alpha^2 T^2}
{|p_4|^4}]}}{p_4^2+\alpha^2}=\raisebox{-2.5mm}{$\buildrel{\displaystyle\lim}
\over{\scriptstyle\alpha\to 0}$}\frac{{\displaystyle exp[-\frac{T^2}
{\alpha^2}]}}{2\alpha}cth[\frac{\beta\alpha}{2}]=0
\end{eqnarray}
and indeed is equal to zero exactly (in any orders of T). However this is not
the case if another prescription (as it was done in Ref.[10]) is used. In this
case
one finds
\begin{eqnarray}
&&\frac{1}{\beta}\sum_{p4}\frac{1}{p_4^2}\longrightarrow\\
&&\hspace{0cm}
\raisebox{-2.5mm}{$\buildrel{\displaystyle\lim}\over{\scriptstyle\alpha\to 0}$}
\frac{1}{\beta}\sum_{p4}\frac{p_4^2}{(p_4^2+\alpha^2)^2}=\raisebox{-2.5mm}
{$\buildrel{\displaystyle\lim}\over{\scriptstyle\alpha\to 0}$}\left\{\frac{1}
{4\alpha}(cth[\frac{\beta\alpha}{2}]-\frac{\beta\alpha}{2}
{\displaystyle \frac{1}{sinh^2[\frac{\beta
\alpha}{2}]}})\right\}=\frac{\beta}{12}\nonumber
\end{eqnarray}
which is the same as (20) only in the high temperature region . This means
that the latter ansatz will generate  additional (the next-to-leading order)
terms in calculations made beyond the high temperature region. So these two
prescriptions are not identical althought in the simplest case (see e. g. [10])
this discrepancy (the additional next-to-leading order term) was eliminated by
the dimensional regularization to obtain the well-known result for the Debye
screening length. Of course, this may be not the case in a more complicated
calculation where these additional terms can reproduce the nonvanishing
expressions and give a different result.

\newpage

\begin{center}
{\bf References}
\end{center}

\renewcommand{\labelenumi}{\arabic{enumi}.)}
\begin{enumerate}

\item{ A.K.Rebhan, Phys. Rev.{\bf D48} (1993) R3967.}

\item{ A.K.Rebhan, in Proc. 3rd Workshop on Thermal Field Theories
and Their Applications, Banff, Canada, Aug. 1993, World Scientific
( F.C. Khanna et al., eds) p.469.}

\item{ R. Baier and O.K.Kalashnikov, Phys.lett.{\bf B328} (1994) 450.}

\item{S. Peign\'e and S.M.H.Wong, Orsay preprint LPTHE-Orsay 94/46,
hep-ph/9406276.}

\item{ A.K.Rebhan, DESY preprint 94-132 (1994), hep-ph/9408262.}

\item{ E.Braaten and A.Nieto, Northwesten University
preprint NUHEP-TH-94-18, hep-ph/9408273.}

\item{ P.V.Landshoff, Phys. Lett.{\bf B169} (1986)69.}

\item{ G. Leibbrandt, Rev. Mod. Phys.{\bf 59} (1987) 1067.}

\item{ P.V.Landshoff, Phys. Lett.{\bf B227} (1989) 427.}

\item{ G. Leibbrandt and M.Staley, in Proc. 3rd Workshop on Thermal Field
Theories and Their Applications, Banff, Canada, Aug. 1993, World Scientific
( F.C. Khanna et al., eds ) p.249.}

\item{ S.M.H.Wong, Orsay preprint LPTHE-Orsay 94/114,hep-ph/9501032. }

\item{ K.Kajantie and J.Kapusta, Ann. Phys.{\bf 160} (1985) 477.}

\item{ U.Heinz, K.Kajantie and T.Toimela, Ann. Phys.{\bf 176} (1987) 218.}

\item{ V.V. Klimov, Sov. J. Nucl. Phys. {\bf 33} (1981) 934;
 Sov. Phys.-JETP {\bf 55} (1982) 199.}

\item{ H.A.Weldon, Phys. Rev.{\bf D26} (1982) 1394. }

\item{ O.K.Kalashnikov, Fortsch. Phys. {\bf 32} (1984) 405.}

\item{ F.Flechsig, A.K.Rebhan and H.Schulz, DESY preprint 95-022 (1995),
ITP-UH-06/95, hep-ph/9502324.}

\end{enumerate}
\end{document}